\pdfoutput=1
\documentclass[conference,10pt]{IEEEtran}
\usepackage [table]{xcolor}
\usepackage{enumitem}
\usepackage{graphicx}
\usepackage{epstopdf}
\usepackage{amssymb}
\usepackage{amsmath,amsthm}
\usepackage{comment}
\usepackage{longtable}
\usepackage{multirow}
\usepackage{booktabs}
\usepackage{pslatex}
\usepackage{tabularx}
\usepackage{microtype}
\usepackage [table]{xcolor}
\usepackage{tikz}
\newcommand*\circled[1]{\tikz[baseline=(char.base)]{ \node[shape=circle,draw,color=white,fill=gray,inner sep=0.5pt] (char){#1};}}
\DeclareRobustCommand*\circledLetter[1]{\tikz[baseline=(char.base)]{ \node[shape=circle,draw,color=white,fill=black,inner sep=0.5pt] (char){#1};}}
\usepackage{etoolbox}
\usepackage{tikzpagenodes}

\usepackage[belowskip=-5pt]{caption}
\usepackage[labelsep=quad,indention=10pt]{subfig}
\usepackage{balance}
\usepackage[font=small]{caption}
\usepackage{url}
\usepackage[bookmarks=false]{hyperref}
\usepackage[noadjust]{cite}
\usepackage{tikz}
\usepackage{pifont}
\newcommand{\cmark}{\ding{51}}%
\newcommand{\xmark}{\ding{55}}%

\usepackage[flushleft]{threeparttable}

\usepackage[ruled,vlined,linesnumbered]{algorithm2e}
\everypar{\looseness=-1}

\def\ie{{i.e.},~}

\usepackage{authblk}
\newcommand{\setanonymous}[1]{
    \newboolean{anonflag}
    \setboolean{anonflag}{#1}
}
\newcommand{\anonymous}[1]{
    \ifthenelse{\boolean{anonflag}}{}{#1}
}
\setanonymous{false}

\title{\LARGE Patient-Driven Privacy Control through Generalized Distillation}
\anonymous{
\setcounter{Maxaffil}{2}
\author[1]{Z. Berkay Celik}
\author[2]{David Lopez-Paz}
\author[1]{Patrick McDaniel}
\affil[1]{SIIS Laboratory, Department of CSE, The Pennsylvania State University\authorcr
{\texttt{\{zbc102, mcdaniel\}@cse.psu.edu}}}
\affil[2]{Facebook AI Research\authorcr
{\texttt{dlp@fb.com}}}
}
\ifthenelse{\boolean{anonflag}}{
    \author{Anonymous\\
    \\
    }}{}
\begin{document}


\maketitle

\begin{tikzpicture}[remember picture,overlay]
    \node[align=center] at ([yshift=1em]current page text area.north) {Accepted to the IEEE Symposium on Privacy-Aware Computing (IEEE PAC) , 2017. Washington DC, USA. };
  \end{tikzpicture}%
  \vspace{-0.2in}

\begin{abstract}
The introduction of data analytics into medicine has changed the nature of patient treatment. In this, patients are asked to disclose personal information such as genetic markers, lifestyle habits, and clinical history. This data is then used by statistical models to predict personalized treatments. However, due to privacy concerns, patients often desire to withhold sensitive information. This self-censorship can impede proper diagnosis and treatment, which may lead to serious health complications and even death over time. In this paper, we present privacy distillation, a mechanism which allows patients to control the type and amount of information they wish to disclose to the healthcare providers for use in statistical models. Meanwhile, it retains the accuracy of models that have access to all patient data under a sufficient but not full set of privacy-relevant information. We validate privacy distillation using a corpus of patients prescribed to warfarin for a personalized dosage. We use a deep neural network to implement privacy distillation for training and making dose predictions. We find that privacy distillation with sufficient privacy-relevant information i) retains accuracy almost as good as having all patient data (only 3\% worse), and ii) is effective at preventing errors that introduce health-related risks (only 3.9\% worse under- or over-prescriptions).
\end{abstract}

\section{Introduction}
Data analytics has introduced a sea change in modern healthcare systems. In this, there are many advances in computation that have the potential to impact society positively on analytically driven personalized medicine~\cite{obama2016united}. This effort has lead to \emph{Precision Medicine Initiative}~\cite{precision-medicine}, announced in 2015. This initiative aims to pioneer data-driven healthcare to treat and prevent disease, based on variability in patient data such as genes, environment, and lifestyle.

Personalized medicine relies on building \emph{statistical models} able to predict personalized treatments, tailored to the characteristics of specific patients. In turn, developing such statistical models call for large amounts of patient data. However, in practice, patients often prefer not to divulge certain kinds of data due to privacy concerns. For instance, social stigma and discrimination concerns of patients suffering from mental disorders and sexually transmitted disease, embarrassing themselves in the presence of their physicians, lack of trust to the medical providers are one of the many reasons~\cite{tourangeau2007sensitive, acquisti2005privacy, appelbaum2003privacy, DHS}. In short, personalized medicine is based on statistical models that call for large bodies of patient data, but patients have different levels of willingness to expose their data\footnote{We note that information withholding that we address in this paper and obscuring (patients lying about their sensitive information) are two different critical problems identified in healthcare. We are currently working on obscurity problem which is algorithmically and conceptually more complex, as it is strictly harder than the withholding problem.}.

To illustrate this dilemma, consider the problem of \emph{personalized warfarin dosing}. Warfarin is an oral anticoagulant prescribed to prevent blood clotting. The \emph{International Warfarin Pharmacogenetics Consortium} (IWPC) has built a statistical model to predict personalized warfarin dose~\cite{international2009estimation}. The IWPC model requires several types of patient data, including genetic markers, clinical history, and demographics. Remarkably, the IWPC model outperforms current standard clinical approaches, which rely on fixed-dose prescriptions. However, a practitioner requires all patients to disclose their complete information to predict treatments. For instance, a patient not filling out a particular medication intake status through a medical history form can not get precise treatment.

The previous example illustrates one main shortcoming of current statistical healthcare models: all patients must provide all of their data, or prediction of treatment is impossible with patients withholding their sensitive data~\cite{saar2007handling}. To amend this problem, data imputation is often applied in healthcare for ``filling the missing'' data using the information provided by public (available) data~\cite{rezvan2015rise}. However, imputation incurs cumbersome endeavor: in warfarin example, imputing medical history of patients is more complicated than predicting the warfarin dose. This may incur biased estimates under many sensitive data if the relationship between missing and available data is intricate~\cite{newgard2015missing}. More importantly, imputation is impossible if the missing data is statistically independent of the available data~\cite{carpenter2008missing}. 

\IEEEpubidadjcol 

\begin{figure*}[t!]
\centering
\includegraphics[width=0.93\textwidth]{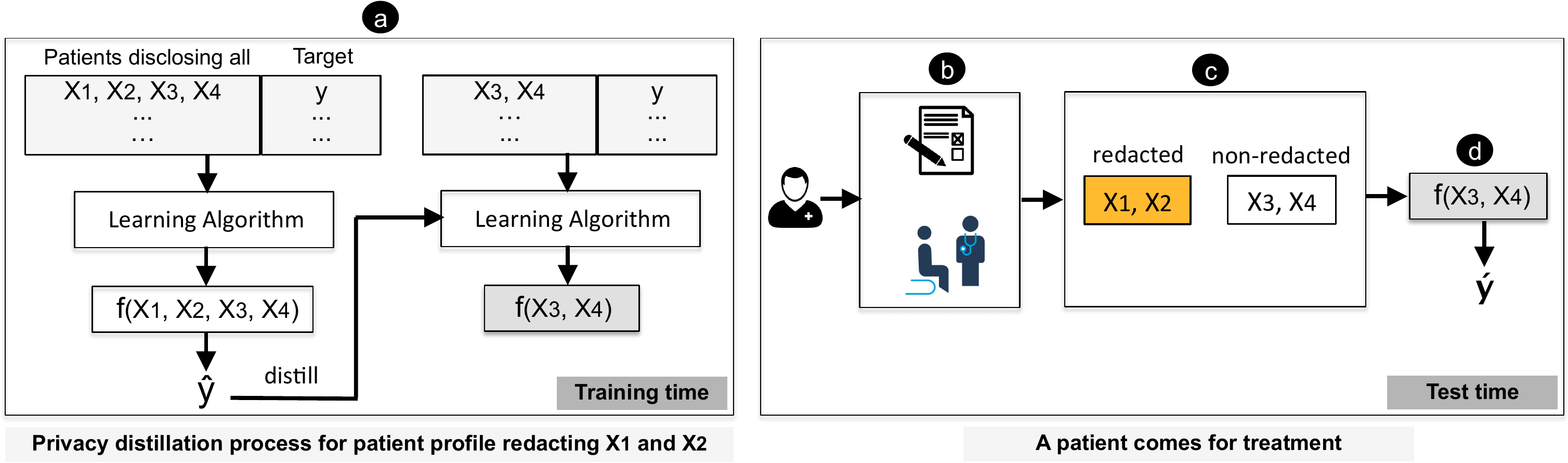}
\caption{Privacy distillation scheme: The distilled model $f(x_3,x_4)$ preserves the privacy of patients at treatment, since it only requires access to their non-redacted data $(x_3, x_4)$. However, the distilled model was trained to imitate the predictions of a full model $f(x_1,x_2,x_3,x_4)$, which had access to the redacted data $(x_1,x_2)$ of other patients and clinical studies. Through this transference of knowledge, the distilled model provides with more accurate treatments than models using only the non-redacted data $(x_3,x_4)$ and treatments \'y are nearly as accurate as models assuming access to the complete patient data ($x_1,x_2,x_3,x_4$).}
\label{fig:distillationFigure}
\end{figure*}

Therefore, data-driven healthcare is in need for a general purpose solution to the problem of making treatments with privacy concerns of patients. We propose one solution to this problem, named privacy distillation.

\vspace{3pt}

\noindent\textbf{Privacy Distillation-} Privacy distillation addresses the privacy concerns of patients for the data used in healthcare statistical models. We define \emph{patient data} that compose features from genetic and clinical data. \emph{Genetic data} include genes or DNA sequences with a known location on a chromosome that its variability plays a major role in treatments. \emph{Clinical data} includes inputs from medication history, psychological profiles, dietary habits, sexual preferences, demographics, and so on. The intuition behind using both types of data is that they are useful for facilitating predictions of the correct treatments. In this setting, we define \emph{redacted features} as those that patients withhold due to privacy concerns.

Privacy distillation works as follows (see Figure~\ref{fig:distillationFigure}): At training time, privacy distillation creates a novel model that retains the accuracy in the absence of redacted features. This is achieved by improving the model training by transferring knowledge from redacted features available across other patients and clinical studies (\circledLetter{a}). At test time, a patient is asked to provide a set of clinical and genetic features for medical treatment through doctor-patient communication, surveys and so on (\circledLetter{b}). The patient chooses to redact features she deems privacy-sensitive  (\circledLetter{c}). The redacted features of a patient are matched with a precomputed \emph{patient profile} that defines different patient disclosure behaviors among patients, and corresponding distilled model then makes the predictions on non-redacted features (\circledLetter{d}). If a patient desires a profile that has not been precomputed, we adapt two solutions (see Section~\ref{sec:distillationFormula}). The goal of this effort is to develop an algorithm such that the accuracy under the redacted patient data is sufficiently close to that of complete data, and any errors in the model should not induce severe medical side effects. We highlight that our approach is a generalizable solution, we here focus its application to personalized medicine.
  
In this paper, we implement privacy distillation using a deep neural network (DNN) and evaluate its effectiveness in the problem of warfarin dose prediction. We compare \emph{privacy distillation} with two baselines. The first baseline is \emph{partially-redacted}, a model built on public (non-redacted) features. The second baseline is the \emph{non-redacted}, which is the current state-of-the-art linear model trained on complete (redacted and non-redacted) patient data. We evaluate privacy distillation by studying the trade-off between the amount of information disclosed by patients and dose accuracy; then we analyze dose errors introducing health-related risks. In this, we make the following contributions:
\begin{itemize}[topsep=0pt, leftmargin=*, noitemsep]
\item We introduce privacy distillation, a mechanism, that allows patients to control their sensitive information used in healthcare statistical models. Privacy distillation offers an accuracy similar to the non-redacted model across various patient profiles; it increases the error of the non-redacted model by only 3\% and outperforms the accuracy of the partially-redacted model by 13.4\%.
\item We show that accuracy loss in healthcare models leads to medical side effects. In the case of warfarin dose predictions, errors cause stroke, embolism, internal bleeding, and even mortality. Privacy distillation provides personalized treatments that are within safety window of a state-of-the-art linear model for 96.1\% of the patients; whereas partially-redacted models lead to 83.2\%. 
\item We improve the accuracy of the current state-of-the-art warfarin dose linear model by 2.7\% with the use of deep neural networks.
\end{itemize}

\section{Technical Preliminaries}
\label{sec:background}
We introduce the essentials of predictive models necessary to understand healthcare statistical models. We then provide an overview of generalized distillation which our privacy distillation mechanism builds on.

\subsection{Healthcare Statistical Models}
\label{sec:predictive-models}
Statistical models are functions $f : \mathcal{X} \to \mathcal{Y}$ that aim at predicting \emph{targets} $y \in \mathcal{Y}$ given some explanatory \emph{features} $x \in \mathcal{X}$. In the following, we consider real-valued targets $y \in \mathbb{R}$ and real vector-valued features $x \in \mathbb{R}^d$. Statistical models are built using a \emph{dataset} containing pairs of features and targets, denoted by $\mathcal{D} = \{(x_i, y_i)\}_{i=1}^n$, and a loss function $\ell : \mathcal{Y} \times \mathcal{Y} \to [0, \infty)$. The loss function penalizes deviations between true targets and predictions. Learning is then searching for the statistical model $f$ minimizing the average loss:
\begin{equation}
\label{eq:obj1}
  \mathcal{L}(\mathcal{D}, f) =  \frac{1}{n} \sum_{i=1}^n \ell(f(x_i), y_i).
\end{equation}
For instance, in simple linear least-squares regression, the statistical model is linear, i.e. $f(x) = \alpha^\top x + \beta$, and the loss function is the squared loss $\ell(f(x),y) = (f(x) - y )^2$. However, to model complex nonlinear relationships between features and targets, we will not be using linear statistical models, but deep neural networks (DNN)~\cite{lecun2015deep}.

In the domain of healthcare, the applications of statistical models are skyrocketing: examples include personalized medicine prescriptions, online crowdsourcing healthcare, disease risk tests, and personality trait tests. In these applications, the features $x$ often correspond to different features describing a patient (medical history, race, weight, etc.), and targets $y$ correspond to a quantity of interest to improve the patient health (optimal dosage of a particular drug, recovery time, depression level, etc.).

When deploying a statistical model in healthcare applications, there are three well-defined stages: i) data collection, ii) feature selection, and iii) model learning. First, data collection refers to the process of gathering relevant diagnostic data from patients, into a dataset of feature-target pairs $\mathcal{D} = \{(x_i, y_i)\}_{i=1}^n$. Second, feature selection refers to the process of removing unnecessary features or attributes from each of the vectors $x_i$. Third, model learning refers to the process of learning a statistical model $f$ from the data $\mathcal{D}$. These three processes are repeated as needed (collecting data, selecting features, learning models) until a sufficient accuracy is achieved. We call the process of learning a statistical model \emph{training time}. We call the process of using a fully trained statistical model to make predictions \emph{test time}. 

\subsection{Generalized Distillation}
\label{sec:gen-distil}
Model \emph{compression} \cite{ba2014deep} or \emph{distillation} \cite{hinton2015distilling} are techniques to reduce the size of statistical models. Model distillation compresses large models $f_\text{large}$ by training a small model $f_{\text{small}}(x)$ that imitates the predictions of the large model $f_{\text{large}}(x)$. Remarkably, model distillation is often able to compress models without incurring any loss in accuracy~\cite{hinton2015distilling}. Put in math, distillation assumes that the large model $f_{\text{large}}$ has been learned by minimizing Equation~\ref{eq:obj1}, and proceeds to learn the small model $f_\text{small}$ by minimizing 
\begin{equation}
\label{eq:obj2}
 (1-\lambda) \mathcal{L}(\{(x_i,y_i)\}_{i=1}^n, f_{\text{small}}) + \lambda \mathcal{L}(\{x_i,s_i\}_{i=1}^n, f_{\text{small}}),
\end{equation}
where $\lambda \in [0,1]$ is an \emph{imitation parameter} trading-off how much does the small model imitate the big model, versus directly learning the data. In \eqref{eq:obj2}, a second dataset is introduced with targets $s_i = f_{\text{large}}(x_i)/T$; these are the \emph{softened predictions} made by the large model. Here, $T > 0$ is a \emph{temperature parameter} scaling the predictions of the large model\footnote{The temperature parameter was recently interpreted as a defense mechanism to adversarial data perturbations~\cite{papernot2015distillation} and as a mechanism to increase the detection accuracy of security sensitive applications~\cite{celik2016building}.}.

Lopez-Paz et al. recently introduced \emph{generalized distillation}, an extension of model distillation used to compress models built on a set of features into models built on a different set of features \cite{lopez2015unifying}. Generalized distillation is one specific instance of \emph{learning using privileged information}~\cite{vapnik2009new, vapnik2015learning, celik2017feature}, a learning paradigm assuming that some of the features used to train a statistical model will not be available at test time. More formally, generalized distillation assumes that a statistical model will be trained on some data $\{(x_i, x^\star_i, y_i)\}_{i=1}^n$, and trained model will be tested on some data $\{x_j\}_{j=n+1}^{n+m}$. Therefore, the set of features $\{x^\star_i\}_{i=1}^n$ is available at training but not at test time. However, these features may contain important information that would lead to statistical models of higher accuracy. For example, consider the case where $x_i$ is the image of a biopsy, $y_i \in \{-1,+1\}$ specifies if the tissue shown in the biopsy shows cancer and $x^\star_i$ is the medical report of an oncologist. It is reasonable to assume that the medical report $x^\star_i$ contains useful information to classify biopsy images $x_i$, but such information will be unavailable at test time.

\emph{Generalized distillation}~\cite{lopez2015unifying} tackles the problem of learning without unavailable data at test time as follows. First, it trains a model $f(x)$ on the feature-target set $\{x^\star_i, y_i\}_{i=1}^n$ by minimizing Equation~\eqref{eq:obj1}. Second, it trains a second model $f(x)$ by minimizing Equation~\eqref{eq:obj2}, where $s_i = f_{\text{large}}(x^\star_i)/T$. Therefore,  it allows developing objectives able to incorporate such sources of information into predictive models, without requiring them at test time. In the sequel, the features not available at test time will correspond to undisclosed private features. Using these features, we will formulate a new variant of generalized distillation in a regression setting through patients' privacy behaviors: we propose to use knowledge extracted from redacted features of a patient across other databases and clinical studies to improve model training when patients redact a different set of privacy-sensitive information.

\begin{figure*}[t!]
\centering
\includegraphics[width=1\textwidth]{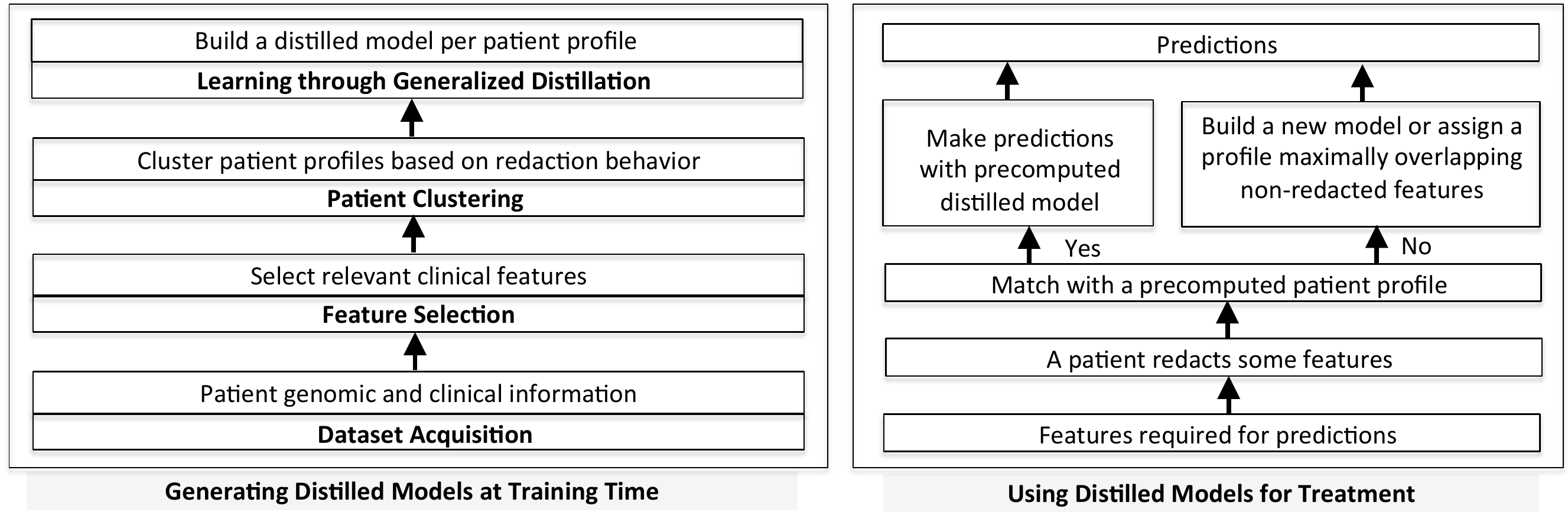}
\caption{Privacy distillation process: (Left) how to learn distilled models based on redaction behaviors, (Right) how to use the distilled model to predict personalized treatments for new patients.}
 \label{fig:privacyDistillationFramework}
 \end{figure*}
 
\section{Patient-driven Privacy}
\label{sec:privacyDistillation}
In the introduction, we stated that patients desire to control data they wish to expose for use in healthcare statistical models. Using the language from Section~\ref{sec:predictive-models}, in order to use a trained model $f(x)$ for treatment, we must have values for all the entries comprising the feature vector $x$. However, in the healthcare domain, the vectors $x$ often describe different types of patient data, and some of these entries may refer to genomic and clinical privacy. Therefore, patients may have different levels of willingness to disclose such data. We give examples of disclosure behavior of the patients in Appendix~\ref{appendix-B}.

To make predictions under redacted inputs, one may consider data imputation to fill the blanks with estimated values. Yet, imputation is impractical because it is either hard to apply under high dimensional redacted (missing) inputs or impossible if redacted inputs are statistically independent of the public (available) inputs. A review of this literature is given in~\cite{chen2012boosting}.

We now start with technical prerequisites required for developing a model that supports patient-driven privacy. We then develop a mechanism for healthcare statistical models complying with the prerequisites, named \emph{privacy distillation}. 

\vspace{2pt}

\noindent\textbf{Patient-driven privacy prerequisites-} We outline the characteristics required for a model to implement an effective and complete patient-driven mechanism:
\begin{enumerate}[leftmargin=*, noitemsep]
\item Feature selection- Given a large pool of patient data; selection of relevant clinical and genomic features is required for a model to improve the accuracy.

\item Control over type and amount of features- The model should consider patients' desire to redact information that is deemed privacy-sensitive.
 
\item Disclosure behavior of patients-  Understanding the disclosure behavior of patients on redacted features is necessary to build efficient models. 

\item Maintaining accuracy- The model under sufficient redacted features should retain the accuracy as almost as good as having complete patient data.
 
\item Patient safety- The prediction errors should not introduce any serious health-related risks.
\end{enumerate}
The rest of this section describes how we address these characteristics with privacy distillation.
 
\subsection{Privacy Distillation}
\label{sec:distillationFormula}  
Privacy distillation is a mechanism to build healthcare statistical models that allow patients to define their own non-redacted and redacted features while providing them with safe and accurate personalized treatments. It comprises two main stages (see Figure~\ref{fig:privacyDistillationFramework}). The first stage trains a statistical model that includes four steps: data acquisition, feature selection, patient clustering and learning via generalized distillation (Section~\ref{sec:distillation-training}). The second stage addresses how to use the distilled model to predict personalized treatments for new patients. It includes two steps: assigning patients to a profile and making predictions (Section~\ref{distillation-testing}).

\vspace{2pt}

\noindent\textbf{Motivating Example-} We introduce the privacy distillation through a simple example. Suppose we wish to treat a patient suffering from Post-Traumatic Stress Disorder (PTSD) with a patient profile ``no sexual assault features''. A patient profile describes a set of patients providing all their features except their background about sexual assault. Privacy distillation proceeds as follows. As a first step, privacy distillation defines a set of patient profiles. Each patient profile describes a different privacy behavior. In the following, let us denote all the patient features by $(x_1, x_2, x_3, x_4)$, and $(x_1,x_2)$ the sexual assault features. Thus, the patient profile ``no sexual assault features'' considers the features $(x_1,x_2)$ as redacted, and the features $(x_3,x_4)$ as non-redacted. Once defined, privacy distillation builds a model for each patient profile. First, privacy distillation  builds a model $\hat{y} = f(x_1,x_2,x_3,x_4)$ that uses all the features available across other patients and clinical studies to estimate the personalized treatment $\hat{y}$. In particular, this model is built to predict the true clinical treatments $y$ available in some database. Second, privacy distillation builds a model $\tilde{y} = f(x_3,x_4)$, this time only on non-redacted features. However, the model $\tilde{y} = f(x_3,x_4)$ is built to learn the true clinical treatments $y$ available in our database, \emph{as well as to imitate the treatments $\hat{y} = f(x_1,x_2,x_3,x_4)$ predicted by the model that uses all the features}. Finally, when a new patient belonging to the ``no sexual assault features'' profile comes for treatment, we can use the model $\tilde{y} = f(x_3,x_4)$ to provide a personalized treatment, while preserving the desired level of privacy. 

\subsubsection{Training Distilled Models}
\label{sec:distillation-training}
This section presents the four steps of training distilled models: data acquisition, feature selection, patient clustering and learning distilled models.

\vspace{2pt}

\noindent\textbf{Data Acquisition-} We collect a set of feature-target pairs $\mathcal{D} = \{(x_i, y_i)\}_{i=1}^n$, where $x_i \in \mathbb{R}^d$ is a vector of $d$ features (both clinical and genetic) describing the $i$-th patient, and $y_i \in \mathbb{R}$ is the target defined for a patient. In our experiments, we use real data collected on patients prescribed to warfarin from multiple medical institutions (see Section~\ref{sec:experimentalsetup}).

\vspace{2pt}

\noindent\textbf{Feature selection-} The raw patient data $\mathcal{D}$ may include multiple irrelevant features by means of questionnaires, health history, lifestyle choices, etc.. These irrelevant features contain no information about the problem of interest, and make the analysis more complicated due to the curses of high-dimensional data~\cite{hund2015analysis}. To alleviate this issue, we preprocess the raw data $\mathcal{D}$ by discarding those features irrelevant to the problem. In particular, we use \emph{Backward Attribute Elimination} (BAE)~\cite{kohavi1997wrappers, Li-etal16} algorithm which is found in previous studies to be highly effective in the dataset we use. The BAE algorithm starts by building $d-1$ statistical models, where the $i$-th statistical model is built using all the features except the $i$-th one, for all $i = 1, \ldots, d$. Next, the BAE algorithm chooses the model with the highest accuracy and decides that the feature that was excluded from that model is irrelevant to the problem. The BAE algorithm proceeds then to build $d-2$ statistical models, in search for the next irrelevant feature. This process is repeated while the desired prediction accuracy is retained. We perform BAE on clinical features since genetic features are often verified by experts. In the following, we denote by $\mathcal{D}$ the dataset obtained after data collection and feature selection.

\vspace{2pt}

\noindent\textbf{Patient Clustering-} We cluster our data $\mathcal{D}$ into $K$ different \emph{patient profiles}. Each patient profile describes a different \emph{disclosure behavior}: what features are understood as public, and what features are understood as privacy-sensitive. The patient profiles are decided by analyzing the different disclosing behavior patterns found in the dataset. The $k$-th patient profile is a subset of $\mathcal{D}$ is defined as follows: 
$$\mathcal{D}^k = \{(x^k_i, x^{k,*}_i, y^k_i)\}_{i=1}^n.$$
In the previous, the vector $(x^k_i, x^{k,*}_i, y^k_i)$ contains the features of the $i$-th patient from the $k$-th patient profile. More specifically, $x^k_i$ are non-redacted features, $x^{k,*}_i$ are redacted features, and $y^k_i$ are targets. We note that we find the disclosure behavior of users by examining the dataset. However, this process can be generalized to a user study to collect their privacy behaviors. This might allow better identification of disclosure behavior of patients; thus the number precomputed patient profiles. We defer conducting a user study to future work.

\vspace{2pt}

\noindent\textbf{Learning Distilled Models-} We now introduce learning distilled models, which is a technique we propose that improves model training under redacted features. Distilled models are adapted from generalized distillation procedure introduced in Section~\ref{sec:gen-distil}, to suit our goal of improving model accuracy in the face of patients redacting private features. 

Our intuition is that not all patients redact the same data; thus, non-redacted data can be acquired from other patients or controlled clinical studies. This is verified through analysis of patient data collected from multiple databases in Section~\ref{sec:experimentalsetup}. Therefore, we transfer the knowledge acquired from non-redacted data of other patients when a patient redacts data.

We tackle the problem of learning with redacted information as follows. First, we train a privileged model $f_\text{priv}(x)$ on the feature-target set $\{x^\star_i, y_i\}_{i=1}^n$. Second, we train a \emph{distilled} model $f_\text{dist}(x)$.  More precisely, given patient profiles, we build one model per patient profile, to obtain $K$ distilled statistical models $f^1,\dots, f^K$. The objective of the $k$-th distilled model is minimized as follows:
\begin{equation}\label{eq:obj3}
 \overbrace{(1-\lambda) \mathcal{L}(\{(x^k_i,y_i)\}_{i=1}^n, f^k)}^\textrm{predict data} + \overbrace{\lambda \mathcal{L}(\{x^k_i,f^{k,*}(x^{k,*}_i)\}_{i=1}^n, f^k)}^\textrm{imitate private model},
\end{equation}
 
where the model $f^{k,*}$ is learned on the set $\{(x^{k,*}_i, y^k_i)\}_{i=1}^{n_k}$. This, we call $f^{k,*}$ a \emph{privileged model}, since it had access to the redacted features at training time. Therefore, the distilled model learns by simultaneously imitating the privileged predictions of the privileged model and learning the targets of the original data. The objective is independent of the learning algorithm; thus it can be minimized using arbitrary models (see Section~\ref{sec:discussion}).

Generalized distillation may have a major problem when used for regression. This is because of the objective formulation through a temperature parameter: $\mathcal{L}(f(x_i),y_i) + \lambda\mathcal{L}(f(x_i),f({x^*}_i)/T)$. Consider a temperature parameter for $T$ = 50, this would be $\mathcal{L}(f(x_i),y_i) + \lambda\mathcal{L}(f(x_i),f({x^*})/50)$ which is almost equal to: $\mathcal{L}(f(x_i),y_i) + \lambda\mathcal{L}(f(x_i),0)$. Using the mean square error, the error is computed: $(f(x_i)-y_i)^2 + \lambda f(x_i)^2$. That is, it penalizes the large predictions, and drives predictions to zero; thus the use of temperature parameter may increase the error of the distilled model while treating patients (\ie underestimates the predictions). Therefore, we formulate the objective~\ref{eq:obj3} with use of imitation parameter $\lambda$ in contrast to having both imitation and temperature parameters. 

In our formulation, the imitation parameter $\lambda$ controls the trade-off between privacy and accuracy. For $\lambda \approx 0$, the objective~\eqref{eq:obj3} approaches the previously introduced standard objective~\eqref{eq:obj1} (Section~\ref{sec:predictive-models}), which amounts to learning a model $f^k$ solely on non-redacted features. However, as $\lambda \to 1$, the objective~\eqref{eq:obj3} transfers the knowledge acquired by the privileged model $f^{*,k}$ into the model $f^k$. The intuition is that whenever privileged model $f^{k,*}$ makes a prediction error at patient $x^k_i$, the model $f^k$ should forget about getting that patient right, and focus on the rest of patients in that patient profile. These \emph{teachings} from the privileged model do, in many cases, significantly help learning process of $f^k$.

To conclude, we balance the privacy and accuracy by using redacted features provided by external patients (\ie not in that profile) when learning the model but ignoring such features when providing treatment to the patients. This is achieved by complementing the training of the model by allowing the transference of knowledge about redacted attributes across patients and clinical studies. As a consequence,  models trained with privacy distillation do not require all data from patients and aims to retain the accuracy under redacted features close to that of complete features.

\subsubsection{Using Distilled Models for Treatment}
\label{distillation-testing}
At test time, a new patient $x_0$ decides what features to disclose, and what features to redact. We then assign the patient $x_0$ to the profile $k$ that maximally overlaps with respect to which features are disclosed (non-redacted features). Then, the distilled model $f^k$ associated with that profile is used to make a prediction. 

If none of the $K$ patient profiles properly describes the disclosure behavior of the patient $x_0$, we can adopt one out of two solutions to maximize the accuracy of her treatment. First, we can build a new patient profile $\mathcal{D}^{k+1}$ and statistical model $f^{k+1}$ which mimics the disclosure behavior of the new patient. This solution would translate into a small computational overhead\footnote{In our experiments; we train 5K patient samples with 65 features on 2.6GHz 2-core Intel i5 processor with 8GB RAM less than one minute including optimal parameter search.} at test time.  Second, we can find an already built profile which discloses every feature that the new patient discloses, but is more conservative about other features. This solution would translate into a potential loss of accuracy since not all of the disclosed features by the new patient would be used for treatment at test time.

\section{Evaluation}
\label{sec:evaluation}
We now evaluate privacy distillation at the problem of warfarin dosing; wherein data is collected from a broad patient population to determine the proper personalized dose. We start with basic clinical information essential to understanding warfarin, as well as review the current state-of-the-art models used to predict its personalized dosage (Section~\ref{sec:modelingwarfarin}). We then describe experimental setup, including the data acquisition, the selection of features, the definition of patient profiles, the baseline models against which we will compare, and the evaluation metrics that we care about (Section~\ref{sec:experimentalsetup}). We evaluate the performance of privacy distillation on the tension between privacy and dose predictions under various patient profiles (Section~\ref{sec:distillation-accuracy}). Finally, we explore the tension between privacy and more tangible utility of patient health safety (Section~\ref{sec:safety}). Our key findings are as follows:
\begin{itemize}
\item  Privacy distillation provides only 3\% less accurate personalized treatments than the state-of-the-art linear model that have access to the complete patient data, and 13.4\% more accurate than the models that ignore privacy-sensitive data.

\item A loss in dose accuracy in healthcare models introduces over-and under-prescriptions that lead to stroke, embolism, and bleeding. Privacy distillation increases the under- and over-prescriptions by 3.9\% of the patients over the state-of-the-art linear model, whereas models constructed without patients' sensitive features (\ie public features) lead to 16.8\% of the patients.

\item By using deep neural networks, we improve the accuracy of the state-of-the-art model by 2.7\%. 
\end{itemize}

\begin{table}[t!]
\centering
\renewcommand{\arraystretch}{1}
\resizebox{\columnwidth}{!}{
\begin{tabular}{p{3.3cm} p{4.7cm}}
\bf{Attribute} & \bf{Patient sample} \\ \midrule
\multicolumn{2}{c}{\textbf{Demographic}} \\
Race  & Black, Black or African American \\
Ethnicity  & Both not Hispanic or Latino \\
Age & 50 - 59 (binned age in years)\\
Gender & male\\
\multicolumn{2}{c}{\textbf{Background (clinical history)}} \\
Height (cm) & 161.29 \\ 
Weight (kg) & 86.1 \\ 
\multirow{3}{*}{Comorbidities}    & no cardiomyopathy,\\
                                  & no hyperlipidemia, \\
                                  & no hypertension \\ 

\multirow{3}{*}{Medications}      & not aspirin,\\
                                  & not simvastatin, \\
                                  & ...\\ 
Macrolide antibiotics & Yes \\
Valve Replacement & No \\ 
Diabetes & Not present \\ 
Herbal medications, & \multirow{2}{*}{Yes} \\
vitamins, supplements & \\
\multicolumn{2}{c}{\textbf{Genotypic}} \\
Cyp2C9 genotypes & *3 \\
VKORC1 SNP rs9923231 & A/G \\
\multicolumn{2}{c}{\textbf{Phenotypic}} \\
Current Smoker & Yes \\ \hline
\end{tabular}}
\caption{Data used for personalized warfarin dose prediction.}
\label{table:patientExample}
\end{table}

\subsection{Warfarin Dose Algorithm}
\label{sec:modelingwarfarin}
Warfarin, known by the brand name Coumadin, is a widely prescribed (over 20 million times each year in the United States) oral anticoagulant used to prevent blood clots from forming mostly in the heart and lungs. Unfortunately, wrong dosages of warfarin may lead to serious and even fatal adverse consequences. On the one hand, under-prescriptions of warfarin will render the treatment useless. On the other hand, over-prescriptions of warfarin may cause hemorrhagic strokes and major bleeding~\cite{abraham2015comparative}. Current practice suggests fixing the initial dose to 5 or 10 milligrams per day~\cite{shepherd2015best}. Then, patients undertake regular blood tests to measure how long it takes for blood clots to form. This measure is referred to as the International Normalized Ratio (INR). The subsequent warfarin dosages are then designed to keep the patient's INR constant and within the desired levels. 

The International Warfarin Pharmacogenetics Consortium (IWPC) performed a study on a large and diverse patient population prescribed to warfarin, for the purpose of determining rules for an accurate personalized warfarin dosage~\cite{international2009estimation}. The study concluded one of the largest and most comprehensive datasets to date for evaluating personalized warfarin. A long line of work culminated on a linear \emph{pharmacogenetic model} that outperformed all previous approaches to personalized warfarin dosage prediction\footnote{The algorithm has been given to doctors and other clinicians for predicting ideal dose of warfarin.}. The model combines genotypic, demographic, phenotypic, and background patient information~\cite{international2009estimation,ramirez2012predicting,kamali2010pharmacogenetics}.
The genotypic information concerns genes VKORC1 and CYP2C9. While the VKORC1 gene encodes the target enzyme of warfarin (vitamin K), the CYP2C9 gene has an effect on warfarin metabolization~\cite{dean2013warfarin}. The demographic and phenotypic information concern various patient characteristics and habits, such as their smoking status. The background information involves the clinical history of a patient, including medications that may interfere their INR. All of this information highlights the variability across patients; thus the necessity of personalized treatment~\cite{ramirez2012predicting}.

\begin{figure}[t!]
\centering
\includegraphics[width=1\columnwidth]{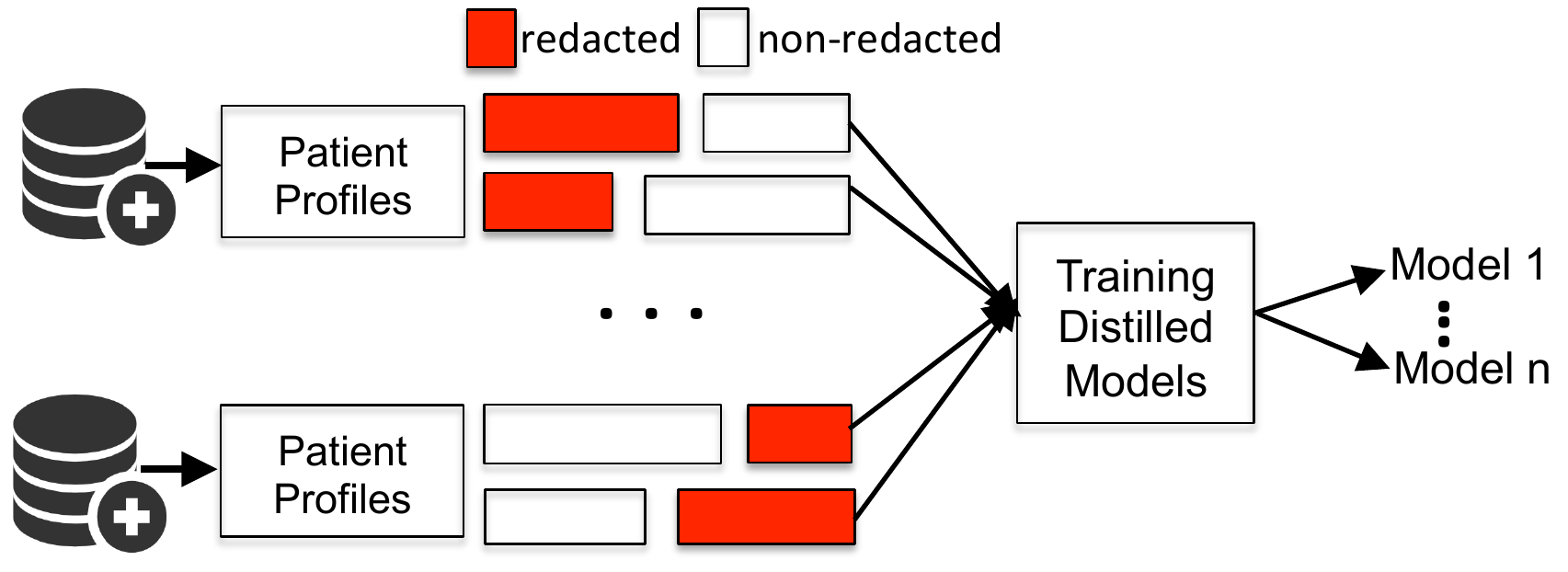}
\caption{Privacy distillation process for privacy behaviors observed in multiple dataset.}
 \label{fig:warfarin-distill}
 \end{figure}

\subsection{Experimental Setup}
\label{sec:experimentalsetup}
In the following, we describe the steps used to set up our experiment for warfarin dose prediction. This section parallels the steps described in Section~\ref{sec:privacyDistillation}.

\vspace{2pt}
\noindent\textbf{Data Acquisition-} We use previously introduced IWPC dataset~\cite{international2009estimation}. The dataset is collected from 21 medical organizations from nine countries and four continents. They are located in Taiwan, Japan, Korea, Singapore, Sweden, Israel, Brazil, United Kingdom, and the United States. The data collection from multiple sources allows to analyze the privacy behavior of patients and to use the non-redacted features in model training required for privacy distillation. For instance, when a patient redacts a set of features, these features are found out to be non-redacted in the dataset of other organizations. This process is illustrated in Figure~\ref{fig:warfarin-distill} and explained throughout this section.

The dataset includes clinical and genotypic data of patients. Table~\ref{table:patientExample} presents an example of features collected from a patient.  We split the clinical data into three categories of six demographic, 24 background, one phenotypic feature. We represent two genotypic features with genetic variants of CYP2C9 as combinations of *1, *2, *3 and VKOCR1 with single-nucleotide rs9923231 of G/G, A/G or A/A genotypes.

 We split the data into two cohorts; the training cohort is used to define patient profiles and to learn the distilled models, and the validation cohort is used to evaluate the accuracy of the distilled models and the errors introducing health-related risks. We convert the categorical variables into numeric factors and standardize features by removing the mean and scaling to unit variance. Overall, the cohorts used in experiments consist of 1221 training and 656 validation samples in which each patient is described by 33 features of four categories, and has warfarin dose as a target.

\vspace{2pt}

\begin{figure}[t!]
\centering
\includegraphics[width=\columnwidth]{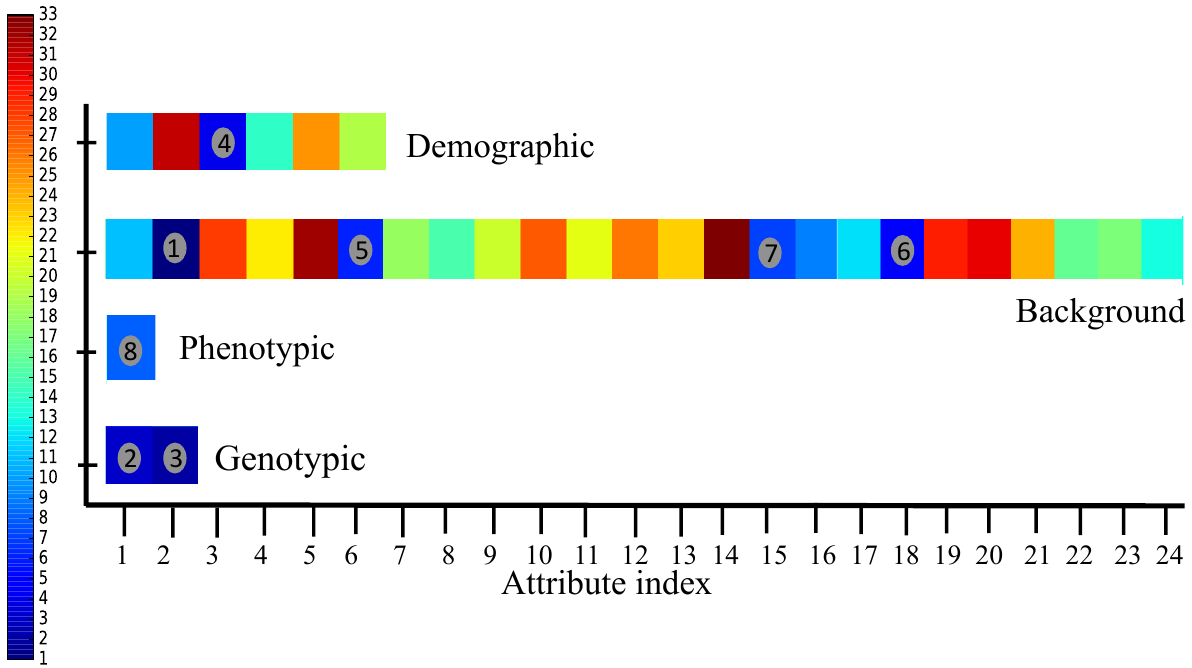}
\caption{An exploration of the features categories and ranking of the each feature. (Refer to the text for feature definitions.) 
}
\label{fig:attributeRankingFigure}
\end{figure}

\noindent\textbf{Feature Selection-} We apply BFE algorithm using least-squares regression (as in reference paper~\cite{international2009estimation}) to infer the relevant features for warfarin dosing. We illustrate the 33 selected features and their categories in Figure~\ref{fig:attributeRankingFigure}. The x-axis shows the feature numbers of each category and y-axis is the name of these categories. The most relevant features are numbered in circles and found to be the primary predictors: Demographic feature of \circled{4} race; genotypic features of \circled{2} Cyp2C9 and \circled{3} VKORC1; background features of \circled{1} weight, \circled{5} use of rifampicin (an antibiotic to treat or prevent a several types of bacterial infections), \circled{6} medications (list of medicines taken), and \circled{7} use of Cordarone (used for treatment of irregular heartbeats), and phenotypic feature of \circled{8} current smoker status.

\vspace{2pt}

\noindent\textbf{Patient Profiles-} We define three patient profiles based on what information patients disclose, with respect to the four categories of demographic, background, phenotypic, and genotypic information (see Table~\ref{table:patient-profiles}). First, \emph{public patients} disclose the information relating to the four categories. Second, \emph{closed patients} disclose the information relating to three out of the four categories. Third, \emph{strict patients} disclose the information relating to only one out of the four categories. For instance, a closed patient refers to ``with all except genotypic'' does not expose genotypic features of VKORC1 and CYP2C9, yet discloses the others, and a strict patient refers to ``genotypic except others'' exposes genotypic features, but nothing else.

\vspace{2pt}

\noindent\textbf{Privacy Distillation Implementation-} We implement privacy distillation using a deep neural network (DNN) with a hidden layer of $32$ rectified linear units each, using the Keras libary~\cite{chollet2015keras} running the Theano backend~\cite{bergstra2010theano}. We use the Adam optimizer to minimize mean squared error loss function~\cite{kingma2014adam}. This setup performance is consistent with more complex architectures that have evaluated with the warfarin dataset and allows us to implement network training as well as the distilled models efficiently. In our experiments, we show the results of privacy distillation with an imitation parameter values of $\lambda \in$ [0,1].

\begin{figure}[h!]
\centering
\includegraphics[width=0.85\columnwidth]{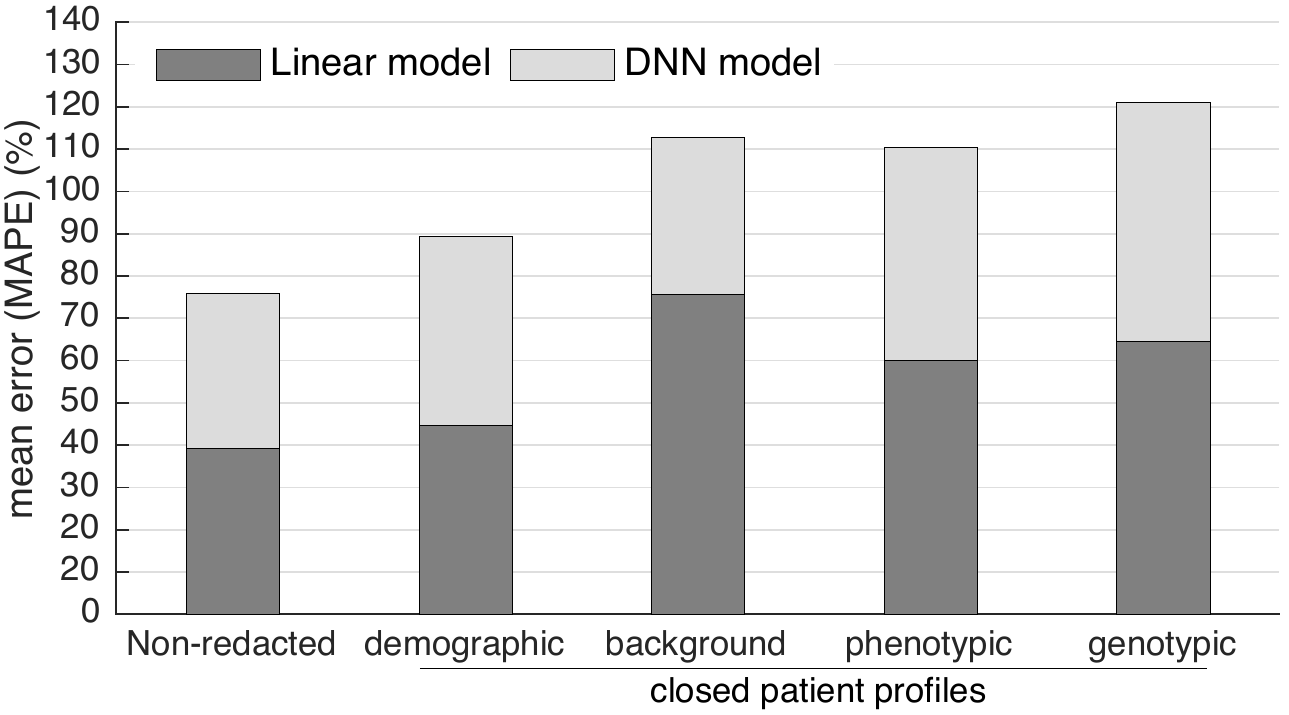}
\caption{Warfarin dose prediction performance measured against linear pharmacogenetic and deep neural network models on a subset of dataset. (Strict patient profiles are similar to closed patient profiles.)}
\label{fig:ols-dnn}
\end{figure}

\vspace{2pt}

\begin{table}[t!]
\centering
\renewcommand{\arraystretch}{1}
\resizebox{\columnwidth}{!}{
\begin{tabular}{llcccc|}
                            &                                              & \multicolumn{4}{c}{Patient data categories}                                                                                           \\\cline{3-6}
                       Legend:     & \multicolumn{1}{l|}{\cmark~ non-redacted, \xmark~ redacted} & \multicolumn{1}{c|}{~\rotatebox{90}{Demographic~}~~} & \multicolumn{1}{c|}{~\rotatebox{90}{Background}~~} & \multicolumn{1}{c|}{~\rotatebox{90}{Phenotypic}~~}& \multicolumn{1}{c|}{\rotatebox{90}{Genotypic}} \\ \hline
\multicolumn{1}{|c|}{Public patient}                       & \multicolumn{1}{l|}{Pharmacogenetic algorithm}                                  & \multicolumn{1}{c|}{\cmark}        & \multicolumn{1}{c|}{\cmark}       & \multicolumn{1}{c|}{\cmark}       & \multicolumn{1}{c|}{\cmark}      \\ \hline
\multicolumn{1}{|l|}{}       & \multicolumn{1}{l|}{With all except demographic}       & \multicolumn{1}{c|}{\xmark}       & \multicolumn{1}{c|}{\cmark}       & \multicolumn{1}{c|}{\cmark}       & \multicolumn{1}{c|}{\cmark}      \\ \cline{2-6} 
\multicolumn{1}{|l|}{Closed} & \multicolumn{1}{l|}{With all except background}        & \multicolumn{1}{c|}{\cmark}        & \multicolumn{1}{c|}{\xmark}      & \multicolumn{1}{c|}{\cmark}       & \multicolumn{1}{c|}{\cmark}      \\ \cline{2-6} 
\multicolumn{1}{|l|}{patient}       & \multicolumn{1}{l|}{With all except phenotypic}        & \multicolumn{1}{c|}{\cmark}        & \multicolumn{1}{c|}{\cmark}       & \multicolumn{1}{c|}{\xmark}      & \multicolumn{1}{c|}{\cmark}      \\ \cline{2-6} 
\multicolumn{1}{|l|}{}       & \multicolumn{1}{l|}{With all except genotypic}         & \multicolumn{1}{c|}{\cmark}        & \multicolumn{1}{c|}{\cmark}       & \multicolumn{1}{c|}{\cmark}       & \multicolumn{1}{c|}{\xmark}     \\ \hline
\multicolumn{1}{|l|}{}       & \multicolumn{1}{l|}{Demographic except others}         & \multicolumn{1}{c|}{\cmark}        & \multicolumn{1}{c|}{\xmark}      & \multicolumn{1}{c|}{\xmark}      & \multicolumn{1}{c|}{\xmark}     \\ \cline{2-6} 
\multicolumn{1}{|l|}{Strict} & \multicolumn{1}{l|}{Background except others}          & \multicolumn{1}{c|}{\xmark}       & \multicolumn{1}{c|}{\cmark}       & \multicolumn{1}{c|}{\xmark}      & \multicolumn{1}{c|}{\xmark}     \\ \cline{2-6} 
\multicolumn{1}{|l|}{patient}       & \multicolumn{1}{l|}{Phenotypic  except others}         & \multicolumn{1}{c|}{\xmark}        & \multicolumn{1}{c|}{\xmark}      & \multicolumn{1}{c|}{\cmark}       & \multicolumn{1}{c|}{\xmark}     \\ \cline{2-6} 
\multicolumn{1}{|l|}{}       & \multicolumn{1}{l|}{Genotypic  except others}          & \multicolumn{1}{c|}{\xmark}        & \multicolumn{1}{c|}{\xmark}      & \multicolumn{1}{c|}{\xmark}      & \multicolumn{1}{c|}{\cmark}      \\  \hline
\end{tabular}
}
\caption{Patient profiles used in experiments.} 
\label{table:patient-profiles}
\end{table}

\begin{figure*}[t!]
\subfloat{%
  \includegraphics[width=\linewidth]{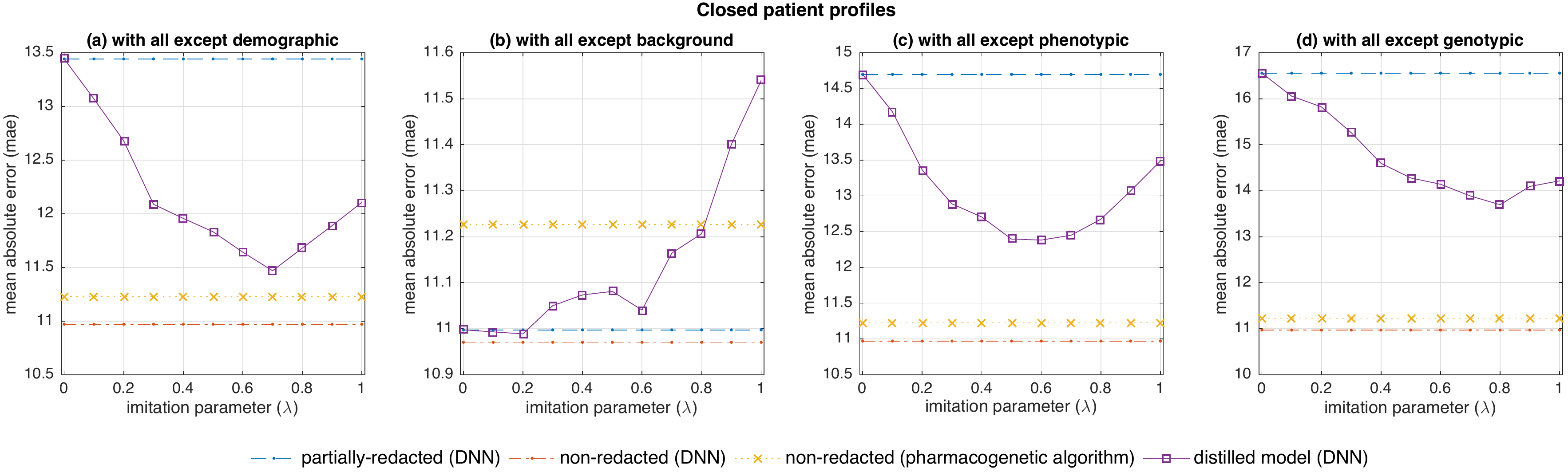}
}

\subfloat{%
  \includegraphics[width=\linewidth]{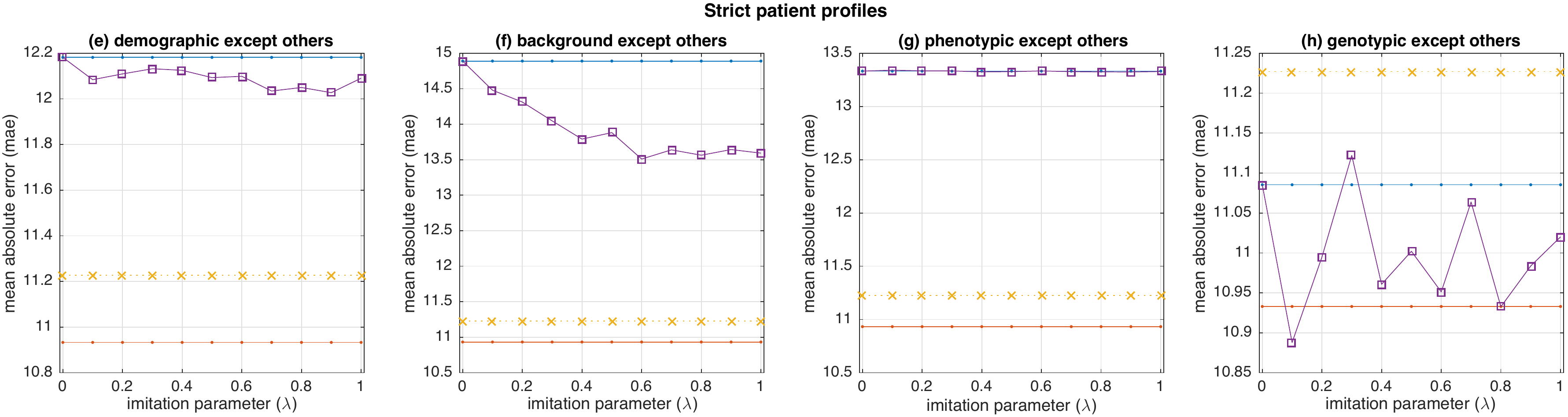}
}
\caption{Comparison of privacy distillation with non-redacted and partially-redacted models. Std. dev. of ten runs falls in on average $\pm$1.7 (see Section~\ref{sec:safety} for medical consequences of models to make dosing decisions).}
\label{fig:PatientOverOneCategory}
\end{figure*}

\noindent\textbf{Evaluation Metrics-} We evaluate the accuracy of models using Mean Absolute Error (MAE) and Mean Absolute Percentage Error (MAPE). These metrics are easily interpreted statistics which is a measure of how far predicted values are away from clinically-deduced warfarin dose and accuracy as a percentage of the error. They are widely used in prediction of real-valued outputs in medical treatments~\cite{international2009estimation}. The lower values of MAE, the closer estimates to the actual dose, indicate better quality of treatment. For each experiment, we report the warfarin dosage MAE and MAPE when averaged over ten random training-validation cohort splits. 

\vspace{2pt}

\noindent\textbf{Models for Comparison-} We evaluate the accuracy of privacy distillation under the patient profiles against two baselines: \emph{Partially-redacted}, which is a model trained only on the information disclosed by the patients,  and \emph{non-redacted} model is trained on the complete patient data. To build a non-redacted model, we reimplement the pharmacogenetic algorithm in Python by direct translation from the authors' implementation~\cite{international2009estimation}. Furthermore, we implement a deep neural network model with the same architecture used in privacy distillation. We find that DNN model gives on average 2.7\% more accurate dose predictions than the current non-redacted linear model. We go on and compare the model performance on partially-redacted models; similar improvements hold both for closed and strict patient profiles (see Figure~\ref{fig:ols-dnn}). To make fair comparisons, we use DNNs to implement the three models: privacy distillation, partially-redacted and non-redacted.

\subsection{Results on Dose Accuracy}
\label{sec:distillation-accuracy}
We evaluate the accuracy of privacy distillation against partially-redacted (dashed line), non-redacted linear pharmacogenetic algorithm (dotted line), and  non-redacted model of DNN (dashed-dotted line). All experiments are performed on previously described patient profiles. Table~\ref{table:perf-results} summarizes the average results of errors and Figure~\ref{fig:PatientOverOneCategory} presents the MAE of privacy distillation with an imitation parameter $\lambda \in$ [0,1]. 

\begin{table}[t!]
\centering
\renewcommand{\arraystretch}{1.15}
\setlength{\tabcolsep}{1pt}
\resizebox{\columnwidth}{!}{%
\begin{tabular}{llcccc}
                             &                             & \multicolumn{2}{l}{\bf{Non-redacted model }}           & \multicolumn{1}{l}{}   & \multicolumn{1}{l}{}  \\
                             &                             & MAE                & MAPE (\%)               & \multicolumn{1}{l}{}   & \multicolumn{1}{l}{}  \\
\multicolumn{1}{l|}{\bf{Public}}  & Pharmacogenetic algorithm   & 11.2               & 39.3                    & \multicolumn{1}{l}{}   & \multicolumn{1}{l}{}  \\
\multicolumn{1}{l|}{\bf{patient}} & Our implementation (DNN)    & 10.9               & 36.6                    & \multicolumn{1}{l}{}   & \multicolumn{1}{l}{}  \\
                             &                             & \multicolumn{2}{c}{\bf{Partially-redacted (DNN)}} & \multicolumn{2}{c}{\bf{Distilled (DNN)}} \\
                             &                             & MAE                & MAPE (\%)               & MAE                    & MAPE (\%)             \\
\multicolumn{1}{l|}{}        & With all except demographic & 13.3               & 44.6                    & 11.4                   & 39.7                  \\
\multicolumn{1}{l|}{\bf{Closed}}  & With all except background  & 11.0               & 37.1                    & 11.0                   & 37.0                  \\
\multicolumn{1}{l|}{\bf{patient}} & With all except phenotypic  & 14.8               & 50.4                    & 12.3                   & 43.5                  \\
\multicolumn{1}{l|}{}        & With all except genotypic   & 16.3               & 56.5                    & 13.5                   & 46.8                  \\
                             &                             &                    &                         &                        &                       \\
\multicolumn{1}{l|}{}        & Demographic except others   & 12.2               & 41.8                    & 12.0                   & 40.8                  \\
\multicolumn{1}{l|}{\bf{Strict}}  & Background except others    & 14.3               & 49.1                    & 13.3                   & 45.5                  \\
\multicolumn{1}{l|}{\bf{patient}} & Phenotypic  except others   & 13.3               & 48.5                    & 13.3                   & 48.5                  \\
\multicolumn{1}{l|}{}        & Genotypic  except others    & 11.1               & 37.3                    & 10.9                   & 36.6                  \\
\end{tabular}%
}
\caption{Prediction performance of models for warfarin dose under patient profiles. We note that partially redacted models implemented with pharmacogenetic algorithm performs on average 13.4\% worse than privacy distillation.}
\label{table:perf-results}
\end{table}

We first evaluate the effectiveness of privacy distillation on closed patient profiles. As shown in Figure~\ref{fig:PatientOverOneCategory}(a-d), \emph{privacy distillation with a carefully chosen $\lambda$ outperforms the partially-redacted model, and achieves a similar accuracy to the non-redacted model.} For instance, illustrated in Figure~\ref{fig:PatientOverOneCategory}(a), ``with all except demographic'' is a profile in which a patient does not disclose her demographic features. Assume for now that warfarin dose prediction is possible only using non-redacted features of patients by using the partially-redacted model at test time. The error yields  MAE of 13.3; However, if a patient provides complete data, the error yields 10.9. This indicates 8\% less accurate dose predictions than the non-redacted model. As we detail in Section~\ref{sec:safety}, this difference significantly increases various medical side effects.

Now, we make warfarin dose predictions by distilled models. Returning to Figure~\ref{fig:PatientOverOneCategory}(a), when a patient redacts the demographic information, the distilled model yields MAE of 11.4. This is exactly 3.1\% mean absolute error less than the non-redacted model compared to 8\% of the partially-redacted model. As seen in Figure~\ref{fig:PatientOverOneCategory}(b-c), the results of other closed patient profiles are similar, and both retain dose errors similar to the non-redacted model, and always better than the partially-redacted model. It is important to note that the impact of using models implemented with DNN over linear models is significant. As shown in Figure~\ref{fig:PatientOverOneCategory}(b), a patient with only partially-redacted features gives better results than the original linear method with all features. In addition, we find that when a patient does not disclose genotypic features, the increase in distilled model errors is more observable than other patient profiles (Figure~\ref{fig:PatientOverOneCategory}(d)) as these features are the key variables for warfarin dose predictions (see Section~\ref{sec:distillation-training}).

A similar analysis on the strict patient profiles confirms the tensions between accuracy and the amount of information disclosed by patients (Figure~\ref{fig:PatientOverOneCategory}(e-h)). The benefit of distilled models similar to the closed patient profiles. However, we make two additional important remarks here. First, patients within the ``phenotypic except other'' would solely disclose their ``smoking status''. This results in a very simplistic model with limited information, and yields higher errors than other patient profiles (Figure~\ref{fig:PatientOverOneCategory}(g)). Second, the importance of genotypic features becomes more apparent when using deep neural networks, instead of the original linear pharmacogenetic algorithm (Figure~\ref{fig:PatientOverOneCategory}(h)). That is, the genetic information of patients better generalizes the nonlinear DNN distilled model with the patients used in our experiments and assigns slightly better dosages than the non-redacted model at $\lambda$=0.1.

Overall, the benefit of privacy-distillation is consistent across all patient profiles. On average, privacy-distillation offers dose predictions 5.7\%  and 3.9\% less accurate than those provided by the non-redacted DNN and state-of-the-art linear pharmacogenetic model. However, it is necessary to study the impact of errors introducing health risks as we discuss next.

\subsection{Results on Patient Health}
\label{sec:safety}
In this section, we ask the essential question: Does making dose predictions with privacy distillation introduce health-related risks? To answer this question, we present a study to evaluate the clinical relevance of dose errors when patients redact features. The study aims at analyzing the dose errors that are inside and outside of the warfarin safety window, and the medical side effects of under- or over prescriptions.

\vspace{2pt}
 
\noindent\textbf{Overview-} We design a study to validate the clinical value of privacy distillation on warfarin dose. The study aims at evaluating the dose errors of partially-redacted, non-redacted models and privacy distillation. We consider weekly dosage errors for each patient because using weekly values eliminates the errors posed by the initial (daily) dose. For instance, a small initial dose error may incur more health risks over time (see discussion for details).  We follow the design of the clinical relevance implemented in~\cite{international2009estimation} which is to date the largest completed warfarin analysis. The authors assess the pharmacogenetic algorithm over clinical and fixed-dose algorithms, yet we measure the model efficacy on making dose predictions. We define two arms to get the weekly dosages:
\begin{enumerate}[leftmargin=*, noitemsep]
\item Pharmacogenetic: Use of a DNN implementation of the pharmacogenetic algorithm with complete patient data. 
\item Private: Use of privacy distillation and partially-redacted models. These models are identical to the pharmacogenetic arm; however, we only replace the pharmacogenetic algorithm with privacy distillation and partially-redacted model under different patient profiles.
\end{enumerate}

Recall that pharmacogenetic algorithm (non-redacted model) is the current best-known algorithm for initial dose prediction over clinical and fixed-dose approach.

  \begin{figure}[t!]
\centering
\includegraphics[width=0.9\columnwidth]{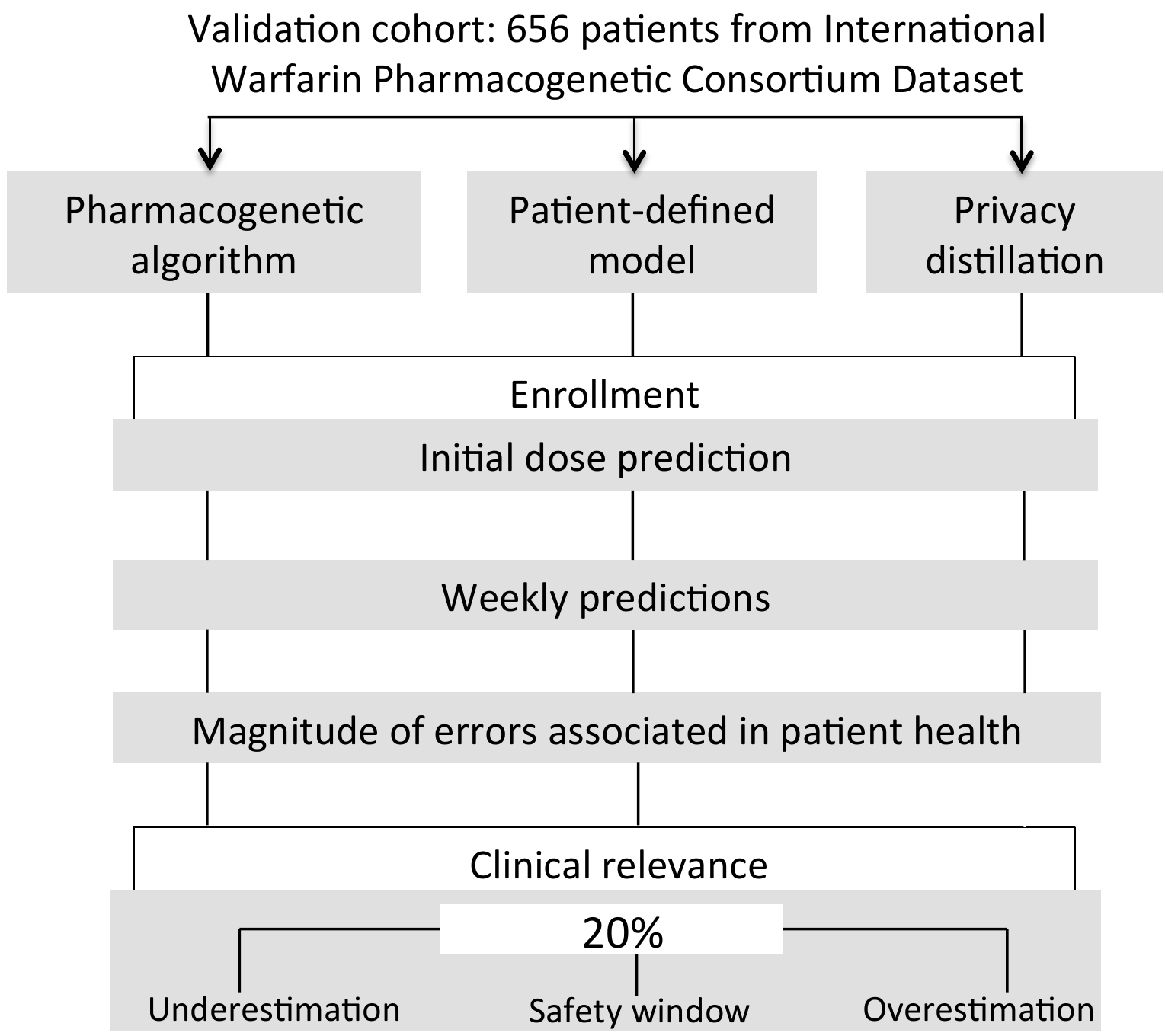}
\caption{Overview of study designed to evaluate the clinical value of privacy distillation. Details of clinical relevance and definition of safety window are given Section~\ref{sec:costSafety}.}
\label{fig:clinicalRelevance}
\end{figure}

\begin{table*}[t!]
\centering
\resizebox{\textwidth}{!}{%
\setlength{\tabcolsep}{12pt}
\renewcommand{\arraystretch}{1}
\begin{tabular}{llcccccc}
                             &                             & \multicolumn{3}{c}{\bf{Non-redacted model}}                           & \multicolumn{1}{l}{} & \multicolumn{2}{c}{\bf{Legend}}                  \\\hline \cline{7-8} 
                             &                             & U (\%)               & SW (\%)              & O (\%)               & \multicolumn{1}{l}{} & \multicolumn{2}{l}{U: Underestimation}      \\
\multicolumn{1}{l|}{\bf{Public}}                       & Pharmacogenetic algorithm   & 21.8                 &~40.9\protect\footnotemark                & 37.3                 & \multicolumn{1}{l}{} & \multicolumn{2}{l}{SW: Safety Window}       \\
\multicolumn{1}{l|}{\bf{patient}}                      & Our implementation (DNN) & 24.9                 & 42.3                 & 33.8                 & \multicolumn{1}{l}{} & \multicolumn{2}{l}{O: Overestimation}       \\
                             &                             & \multicolumn{1}{l}{} & \multicolumn{1}{l}{} & \multicolumn{1}{l}{} & \multicolumn{1}{l}{} & \multicolumn{1}{l}{} & \multicolumn{1}{l}{} \\
                             &                             & \multicolumn{3}{c}{\bf{Partially-redacted model (DNN)}}                 & \multicolumn{3}{c}{\bf{Privacy distillation (DNN)}                    } \\ 
                             &                             & U (\%)               & SW (\%)              & O (\%)               & U (\%)               & SW (\%)              & O (\%)               \\
\multicolumn{1}{l|}{}        & With all except demographic & 28.2                 & 36.1                 & 35.7                 & 22.8                 & 40.4                 & 36.8                 \\
\multicolumn{1}{l|}{\bf{Closed}}  & With all except background  & 23.5                 & 42.0                 & 34.5                 & 23.4                 & 42.3                 & 34.3                 \\
\multicolumn{1}{l|}{\bf{patient}} & With all except phenotypic  & 30.9                 & 33.0                 & 36.1                 & 23.3                 & 36.5                 & 40.2                 \\
\multicolumn{1}{l|}{}        & With all except genotypic   & 31.3                 & 30.1                 & 38.6                 & 27.7                & 33.2                 & 39.1                 \\
                             &                             & \multicolumn{1}{l}{} & \multicolumn{1}{l}{} & \multicolumn{1}{l}{} & \multicolumn{1}{l}{} & \multicolumn{1}{l}{} & \multicolumn{1}{l}{} \\
\multicolumn{1}{l|}{}        & Demographic except others   & 24.0                 & 37.4                 & 38.6                 & 25.1                 & 37.8                 & 37.1                 \\
\multicolumn{1}{l|}{\bf{Strict}}  & Background except others    & 28.9                 & 32.4                 & 38.7                 & 27.5                 & 34.6                 & 37.9                 \\
\multicolumn{1}{l|}{\bf{patient}} & Phenotypic  except others   & 26.8                 & 31.0                 & 42.2                 & 26.2                 & 31.7                 & 42.1                 \\
\multicolumn{1}{l|}{}        & Genotypic  except others    & 20.5                 & 39.2                 & 40.3                 & 20.1                 & 39.8                 & 40.1   \\             
\end{tabular}%
}
\caption{Percentage of patients in validation cohort that are within the safety window, underestimated or overestimated against weekly dosages of clinically-deduced dosages. Std. dev. of ten independent runs falls in on average $\pm$2.5. Partially redacted models implemented with pharmacogenetic algorithm perform on average 24.1\% of the patients in the safety window.}
\label{table:patientSafetyResults}
\end{table*}

\subsubsection{Patient Safety}
\label{sec:costSafety}
To evaluate the impact of errors that introduce health-related risks, we draw patient samples from validation cohort and assign them weekly warfarin dosages with the use of pharmacogenetic and privacy arms (see Figure~\ref{fig:clinicalRelevance}). We then measure the errors between the estimated dose and clinically-deduced ground truths. These errors define how good arms are on the prediction of dosage for a particular patient. The direction of the error comes with a positive or negative sign that points out the under or over-prescribes. We use this value for finding the side effects as detailed next. 

\vspace{2pt}

\noindent\textbf{Safety window-} We calculate the percentage of patients whose predicted dose of warfarin is within a therapeutic safety window. The weekly dose is accepted in the safety window if an estimated dose falls within 20\% of its corresponding clinically-deduced ground truth. This is because the value defines the difference of 1 mg per day relative to the fixed starting dose of 5 mg per day which is often accepted as clinically relevant~\cite{international2009estimation, kimmel2013pharmacogenetic}. The deviations of both directions fall outside of the safety window. Specifically, any dose prediction is an overestimation that is at least 20\% higher than the actual dose and is an underestimation that is at least 20\% lower than the actual dose.

\vspace{2pt}

\noindent\textbf{Health risks-} To identify the side effects of warfarin, we follow  U.S. Food and Drug Administration (FDA) warfarin medical guide for the most common side effects of overestimation and underestimation~\cite{fdaWarfarinRisk}. Taking high-dose warfarin causes INR to be high (thin blood) which causes higher risks of having intracranial and extracranial bleeding. Taking low doses causes INR to be low (thick blood), and warfarin does not protect patients from developing a blood clot, and it causes embolism and stroke. In both cases, extremely high or low doses may lead to death. We note that certain amount of dosing errors, in reality, do not immediately cause these adverse effects, as factors such as commonly used medications and foods with vitamin K also play a role in blood clotting.
 
\subsubsection{Clinical Relevance}
We now evaluate the health-related risks of making dose decisions under patient profiles with partially-redacted, non-redacted and privacy distillation models. Table~\ref{table:patientSafetyResults} shows the percentage of patients within the dose safety window, under-prescriptions, and over-prescriptions. 

Remarkably, \emph{privacy-distillation yields nearly-optimal weekly dose errors, when compared to non-redacted DNN implementation of a pharmacogenetic algorithm that assumes access to complete patient information}. It yields on average 5.3\% fewer patients in the safety window. For instance, ``with all except demographic'' closed patient profile gives 22.8\% under-prescribes, 40.4\% in the safety window and 36.8\% over-prescribes. This value is only 1.9\% less than the non-redacted model and 4.3\% more than the results of the partially-redacted model. It is interesting to report that the direction of the error changes. The percentage of overestimated patients is increased from 33.8\% to 36.8\% and underestimated patients is decreased from 24.9\% to 22.8\%.

Analyzing the results more in detail, we find that some patient profiles yield a lower percentage of patients in the safety window compared to other profiles. ``With all except genotypic'' and ``Phenotypic except others'' patient profiles yield 33.2\% and 31.7\% of the patients in the safety window compared to the 42.3\% of public patients. As discussed in Section~\ref{sec:distillation-accuracy}, the high amount of information disclosed by patients such as using only phenotypic feature or disclosure of important variables such as genotypic features causes loss of information in model learning. Thus, over- and under-prescriptions become more apparent than other patient profiles. However, privacy distillation always performs better treatments than those of partially-redacted models.

To conclude, privacy distillation gives nearly optimal dosage error across all patient profiles. It offers 5.3\%  and 3.9\% fewer patients in the safety window than those provided by the non-redacted DNN and state-of-the-art linear pharmacogenetic model. In turn, it is effective at preventing errors that introduce health-related risks.

\addtocounter{footnote}{0}
\footnotetext{Note: We remark that our results are similar to those found in reference paper~\cite{international2009estimation}. These results represent the current best-known pharmacogenetic method for predicting initial warfarin dose, and often gives more accurate dose prescriptions than current clinical practices as difficulty in establishing an initial dose of warfarin varies by a factor of 10 among patients~\cite{international2009estimation,kasner2016warfarin}.}

\section{Discussion}
\label{sec:discussion}
We presented privacy distillation in a regression setting. However, its objective can be formulated to other statistical models such as supervised, semi-supervised, transfer and universum learning~\cite{lopez2015unifying}. In these, two important factors need to be addressed to maximize the accuracy: Model selection and parameter tuning. For the former, the objective function can be minimized using \emph{arbitrary} models, as privacy distillation is a model free mechanism. For the latter, while minimizing the objective, the imitation parameter $\lambda$ should be chosen carefully for finding the optimal distilled model.

It is important to note that our study on predicting warfarin dosages makes the evaluation of privacy distillation different than those solely based on raw accuracy. However, there exist more comprehensive studies such as clinical trials to evaluate long-term health risks. For instance, it is common to observe warfarin dose by subsequently titrating to 90 days by using pharmacokinetic/pharmacodynamics (PK/PD) models, as dose titration after one-week period may change the dosages and pose varying health risks~\cite{anderson2007randomized}. However, long-term studies require observation of INR responses because INR levels are not stable during this period. 

The other crucial point for an evaluation of errors is the models for risk estimation. There exist advanced models to estimate an adverse event occurring in a specific period or as a result of a specific situation. For instance, varying INR levels of a patient can be used to estimate the correlation between stroke and bleeding event~\cite{sorensen2009cost}. These models define more detailed health risks through various measures of the blood clot. However, the input/output of the model components interacts with detailed clinical patient information for precise estimation. Privacy distillation can be easily integrated with more exhaustive clinical trials under these models. 

\section{Related Work}

\noindent\textbf{Privacy-sensitive Information-} Clarke have categorized the types of privacy and outlined specific protections~\cite{clarke1988information}. The categories include the privacy of the person, privacy of personal behavior, the privacy of personal data, and privacy of personal communication. Pertaining to these categories, recent studies have analyzed the individuals' behavior of information disclosure both in online and offline contexts\cite{appari2010information,naveed2015privacy}. The common point of the studies is individuals' context-dependent preferences and the subjectivity on the personal matters. We solve these problems with one of our implementation goal of making predictions available under varying number of redacted inputs of patients (\ie patient profiles).
\vspace{1pt}

\noindent \textbf{The Law and Medical Privacy-} The goal of medical privacy is keeping information about a patient confidential. This involves both conversational discretion and medical record security. The Health Insurance Portability and Accountability Act (HIPAA) privacy and security rules~\cite{act1996health,centers2003hipaa} is the baseline law that protects patient medical information. There are also recent laws dealing with genetic information privacy~\cite{hudson2008keeping}. The goal of these regulations is to establish the rights individuals have concerning their health information. Privacy distillation formalizes this goal and addresses the patient control over privacy-sensitive information under these regulations.
\vspace{1pt}

\noindent\textbf{Healthcare Statistical Models-} In recent years, researchers have extensively investigated healthcare models. Futoma et al. used random forests and DNNs for predicting early hospital readmissions~\cite{futoma2015comparison}. Volm et al. developed \emph{in vitro} tests to predict tumors' drug responsiveness for cancer treatment~\cite{volm2015prediction}. Other researchers in pharmacogenomics predicted the dose of the medicines and their responses on patients~\cite{arranz2016pharmacogenetics, biernacka2015international}. These models can be equipped with privacy distillation when a set of features is identified as privacy-sensitive. In addition, privacy distillation handles all stages of modeling, including feature selection and redaction behavior of patients.
\vspace{1pt}

\noindent\textbf{Privacy Threats to Patient Information-} Privacy threats target obtaining sensitive patient information from models or databases. Homer et al. used the public allele frequencies to infer the possibility of the participation of an individual in a genotype database~\cite{homer2008resolving}. Wang et al. showed that statistics on genetics and diseases could be used to identify individuals~\cite{wang2009learning}. Fredrikson et. al. introduced a model inversion attack and used same warfarin dataset to predict a patient's genetic markers from some clinical data and warfarin dosages~\cite{fredrikson2014privacy}. These approaches consider different utility/privacy metrics, as they assume that patients disclose all of their information and attempt to infer sensitive information by studying the relationships between inputs and outputs of the model. 

\vspace{1pt}
 
\noindent\textbf{Defense of Privacy Leaks-} As a response to the privacy threats, researchers suggest prevention and remedies to such breaches~\cite{mclaren2016privacy, celik2017curie, dankar2012application}. These approaches are limited to protecting the privacy of users on inputs required for training or classification. We view our efforts in this paper to be complementary to much these. Privacy distillation can be easily integrated as a user-driven mechanism to strike a balance between accuracy and redacted inputs of users.

\section{Conclusion}
We proposed a patient-driven privacy mechanism named privacy distillation. It is a learning meta-algorithm to construct accurate healthcare statistical models that allow patients to control their privacy-sensitive data. We evaluated the effectiveness of privacy distillation on pharmacogenetic modeling of personalized warfarin dose. In our experiments, privacy distillation outperformed the state-of-the-art healthcare statistical models that ignore privacy-sensitive data. By reusing knowledge about privacy-sensitive information across patients, privacy distillation showed nearly optimal warfarin prescriptions, competing with the idealized models that assume access to complete patient data. We analyzed the impact of the accuracy gains provided by privacy distillation on patient health; this showed a significant reduction in warfarin under- and over-prescriptions. Such accurate predictions can translate into less health-related risks of embolism, strokes, or bleeding. 

This work is the first effort at developing models under redacted users inputs. The capacity afforded by this approach will allow us to make accurate predictions in a wide array of applications requiring private inputs, such as the ones found in medicine, law, forensics, and social networks. In the future, we will explore a wide range of environments and evaluate the ability of privacy distillation to promote its effectiveness.

{\anonymous{
\section{Acknowledgment}
Research was sponsored by the Army Research Laboratory and was accomplished under Cooperative Agreement Number W911NF-13-2-0045 (ARL Cyber Security CRA). The views and conclusions contained in this document are those of the authors and should not be interpreted as representing the official policies, either expressed or implied, of the Army Research Laboratory or the U.S. Government. The U.S. Government is authorized to reproduce and distribute reprints for Government purposes notwithstanding any copyright notation here on.
}}

{\footnotesize{
\bibliographystyle{abbrv}
\bibliography{patientPrivacy} }}

\appendix
\setcounter{table}{0}
\setcounter{figure}{0}
\subsection{ Privacy Concerns of Patients}
\label{appendix-B}
We provide some examples of patients redacting their data due to the privacy concerns. We refer the reader to Koontz~\cite{koontz2013information} for a more comprehensive discussion of privacy concerns of patients and their reasons.

\begin{itemize}[leftmargin=*, noitemsep]
\item U.S. Department of Health \& Human Services (HHS) estimated that 2M Americans did not seek treatment for mental illness~\cite{DHS}.
\item Millions of young Americans that suffers from sexually transmitted diseases do not seek treatment~\cite{DHS}.
\item HHS estimated that 586K Americans did not seek earlier cancer treatment~\cite{DHS}.
\item The Rand Corporation found that 150K soldiers suffering from Post-Traumatic Stress Disorder (PTSD) do not seek treatment~\cite{holdeman2009invisible}. Privacy concerns contribute to the highest rate of suicide among active duty soldiers in 30 years.
\end{itemize}

\end{document}